\documentclass[conference]{IEEEtran}%
\IEEEoverridecommandlockouts

\usepackage{amsfonts}
\usepackage{amsmath}
\usepackage{amssymb}
\usepackage{balance}
\usepackage{ifpdf}
\usepackage{color}
\usepackage[pdftex]{graphicx}

\ifCLASSINFOpdf
\graphicspath{{./pdf/}}
   \DeclareGraphicsExtensions{{.pdf}}
\else
\graphicspath{{./eps/}}
   \DeclareGraphicsExtensions{{.eps}}
\fi

\begin{document}

\title{Distributed Estimation of a Parametric Field \\ Using Sparse Noisy Data}

\author{Natalia~A.~Schmid,~\IEEEmembership{Member,~IEEE,}
        Marwan~Alkhweldi,~\IEEEmembership{}
        and~~Matthew~C.~Valenti,~\IEEEmembership{Senior~Member,~IEEE}
\vspace{-0.4cm}
\thanks{N. A. Schmid, M. Alkhweldi, and M. C. Valenti are with the Department
of Computer Science and Electrical Engineering, West Virginia University, Morgantown,
WV, 26506 USA e-mail: Natalia.Schmid@mail.wvu.edu, malkhwel@mix.wvu.edu, and Matthew.Valenti@mail.wvu.edu.}
\thanks{This work was sponsored by the Office of Naval Research under Award No. N00014-09-1-1189.}
}

\maketitle

\thispagestyle{empty}

\begin{abstract}
The problem of distributed estimation of a parametric physical field is stated as a maximum likelihood estimation problem. Sensor observations are distorted by additive white Gaussian noise. Prior to data transmission, each sensor quantizes its observation to $M$ levels. The quantized data are then communicated over parallel additive white Gaussian channels to a fusion center for a joint estimation. An iterative expectation-maximization (EM) algorithm to estimate the unknown parameter is formulated, and its linearized version is adopted for numerical analysis. The numerical examples are provided for the case of the field modeled as a Gaussian bell. The dependence of the integrated mean-square error on the number of quantization levels, the number of sensors in the network and the SNR in observation and transmission channels is analyzed.
\end{abstract}

\begin{IEEEkeywords}
Distributed estimation, expectation-maximization algorithm, maximum likelihood estimation, distributed sensor network, sparse data
\end{IEEEkeywords}

\section{Introduction}
\label{sec:intro}

Distributed sensor networks provide a platform for many military and civilian applications. Examples include surveillance, monitoring wildlife, or controlling the power grid. Sensor networks built for these applications are intended to solve various problems such as detecting, tracking, classifying, counting, and estimating. They are also required to adhere to a number of physical constraints including power, bandwidth, latency, and complexity. Much research has been reported on each of these topics over the past two decades. In the field of distributed estimation, for example, various estimation problems have been formulated and solved.
Many works choose either to optimize a distributed sensor network with respect to energy consumption during transmission \cite{Li09,Wu08} or impose bandwidth constraints and thus focus on designing an optimal quantization strategy for the distributed network \cite{Ribeiro06}, \cite{Niu2006}. There are few that involve both constraints (see \cite{Cui07} as an example).

Among research groups working on the problem of distributed estimation, there are a few dealing with distributed estimation of a field (a multidimensional function, in general)  \cite{Schabus11,Wang08,Nowak03}. Since in many real-world applications distributed estimation of a multidimensional function may provide additional information that aids in making a high-fidelity decision or in solving another inference problem, we contribute to this topic by formulating and solving the problem of a parametric field estimation from sparse noisy sensor measurements. Distributed target localization is another active area of research \cite{Niu2006}. An iterative solution to this problem is a second contribution of our work.

In this paper, the problem of distributed estimation of a physical field from sensory data collected by a homogeneous sensor network is stated as a maximum likelihood estimation problem. The physical field is a deterministic function and has a known spatial distribution parameterized by a set of unknown parameters, such as the location of an object generating the field and the strength of the field in the region occupied by the sensors. Sensor observations are distorted by additive white Gaussian noise. Prior to transmission, each sensor quantizes its observation to $M$ levels. The quantized data are then communicated over parallel additive white Gaussian channels to the fusion center where the unknown parameters of the underlying physical field are estimated.  An iterative expectation-maximization (EM) algorithm to estimate the unknown parameter is formulated, and a simplified numerical solution involving additional approximations is developed. The numerical examples illustrate the developed approach for the case of the field modeled as a Gaussian bell.

The remainder of the paper is organized as follows. Sec. \ref{sec:Problem Statement} formulates the problem. Sec. \ref{sec:Iterative Solution} develops an EM solution. Sec. \ref{sec:Numerical_Analysis} provides numerical performance evaluation. The summary of the developed results is provided in Sec. \ref{sec:Summary}.

\section{Problem Statement}
\label{sec:Problem Statement}

Consider a distributed network of homogeneous sensors monitoring the environment for the presence of a substance or an object. Assume that each substance or object is characterized by a location parameter and by a spatially distributed physical field generated by it. As an example, a ferromagnetic object can be viewed as a single or a collection of dipoles characterized by a magnetic field that they generate. This field can be sensed by a network of magnetometers placed in the vicinity of the object. Depending on the design of the magnetometers, they may take measurements of a directional complex valued magnetic field or of the magnitude of the field only. The field generated by a dipole decays as a function of the inverse cube of the distance to the dipole. The sensor network does not know a priori the location of the dipole as well as the type and size of the object. However, the type and size of an object can be associated with the strength of the magnetic field. Examples of other physical fields include (1) a radioactive field that can be modeled as a stationary spatially distributed Poisson field with a two-dimensional intensity function decaying according to the inverse-square law or (2) a distribution of pollution or chemical fumes that, if stationary, can often be modeled as a Gaussian bell.


Consider a network of $K$ sensors distributed over an area $A.$ The network is calibrated in the sense that the relative locations of the sensors are known. Sensors act independently of one another and take noisy measurements of a physical field $G(x,y).$ A sample of $G(x,y)$ at a location $(x_k,y_k)$ is denoted as $G_k=G(x_k,y_k).$ The parametric field $G(x,y)$ is characterized by $L$ unknown parameters $\mathbf{\theta}=[\theta_1,\ldots,\theta_L]^T.$ To emphasize this dependence we will use both $G_k$ and $G(x_k,y_k:\theta)$ throughout the text. The sensor noise, denote it by $W_k,$ $k=1,\ldots,K$ is known and modeled as Gaussian distributed with mean zero and variance $\sigma^2.$ The noise of sensors is independent and identically distributed. Let $R_k,$ $k=1,\ldots,K$ be the noisy samples of the field at the location of distributed sensors.  Then  $R_k$ is modeled as $R_k=G(x_k,y_k)+W_k.$


\begin{figure}[!t]
\centering
\includegraphics[width=3.5in]{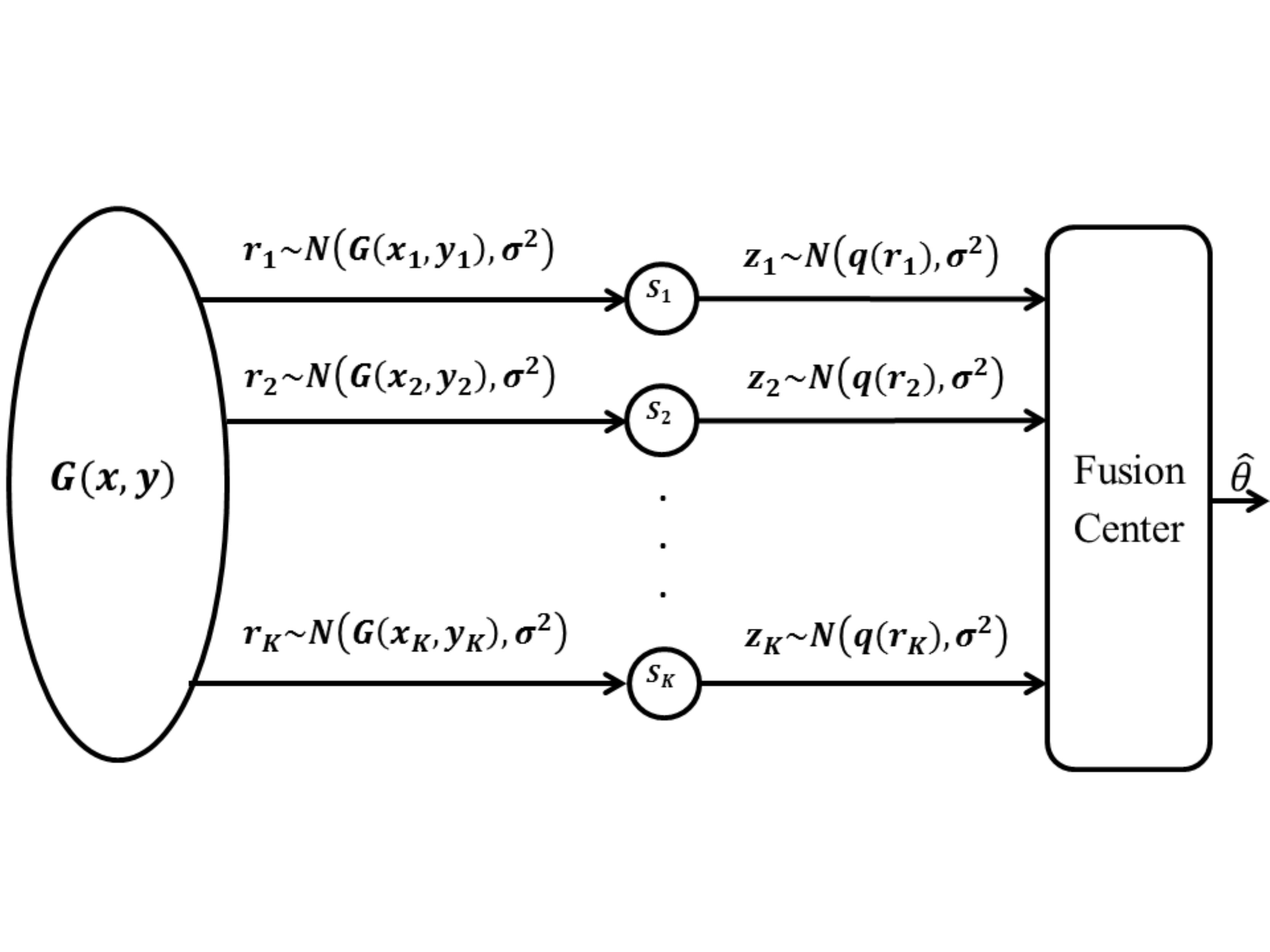}
\vspace{-1.5cm}
\caption{Block-diagram of the distributed sensor network. }
\label{fig:block_diagram}
\end{figure}
Due to constraints that are imposed by practical technology, each sensor may be required to quantize its measurements prior to transmitting them to the fusion center (FC). Assume that a deterministic quantizer with $M$ quantization levels is involved. Let  $\nu_1,\nu_2,\ldots,\nu_M$  be known reproduction points of the quantizer. Denote by $q(R_k)=q_k$ the quantized version of the measurement by the $k$-th sensor. These data are modulated using a digital modulation scheme and then transmitted to the FC over noisy parallel channels. The noise in channels is due to quantization error and channel impairments, denote it by $\tilde{N}_k,$ $k=1,\dots,K.$ Denote by $m(\cdot)$ a modulation function and by $d(\cdot)$ a demodulation function. Let $Z_1,\ldots,Z_K,$ be noisy observations received by the FC. Then each $Z_k$ is given by $Z_k=d(m(q_k))+\tilde{N}_k, \ \  k=1,\ldots,K.$ In this work we assume that $m(\cdot)$ and $d(\cdot)$ are linear and that the demodulator recovers the quantized signal by using a soft thresholding rule. These assumptions allow $Z_k$ be approximated by its asymptotic counterpart $Z_k=q_k+N_k,$ where $N_k$ is a white Gaussian noise with variance $\eta^2.$

Given the noisy measurements and the relative location of the sensors in the network, the task of the FC is to estimate the vector parameter $\mathbf{\theta}.$ A block diagram of the distributed network used for estimation of parameters of a physical field is shown in Fig. \ref{fig:block_diagram}.

In this work we adopt a maximum likelihood (ML) estimation approach to solve the problem of distributed parameter estimation. The joint likelihood function of the independent quantized noisy measurements $Z_1, Z_2, \ldots, Z_K$ can be written as
\begin{eqnarray}
l(\mathbf{Z}) & = & \sum^{K}_{k =1} \log \left(\sum^{M}_{j =1}\ p_{k,j} \exp \left( - \frac{ \left( Z_k - \nu_j \right)^2 }{2\eta^2 } \right)\right),
\label{eq:incomplete_likelihood}
\end{eqnarray}
where $\mathbf{Z}$ is the vector of measurements $[Z_1,Z_2,\ldots,Z_K]^T,$ $p_{k,j}$ are the probabilities for the output of the sensor $k$ to be mapped to the $j$-th reproduction point during the encoding process
\[ p_{k,j} =\int_{\tau_j}^{\tau_{j+1}} \frac{1}{\sqrt{2\pi\sigma^2} } \exp \left( -\frac{ \left( t -G_k \right)^2 }{2\sigma^2 } \right)dt,\]
$\tau_j$ and $\tau_{j+1},$ $j=1,\ldots,M$ are the boundaries of the $j$-th quantization region. The ML solution $\mathbf{\hat{\theta}}$ is the solution that maximizes the expression (\ref{eq:incomplete_likelihood}). For a numerical example in Sec. \ref{sec:Numerical_Analysis}, the field is modeled as a Gaussian bell with three unknown parameters: the strength of the field $\mu$ and the location parameter $(x_c,y_c).$


\section{Iterative Solution}
\label{sec:Iterative Solution}

Since the expression for the log-likelihood function (\ref{eq:incomplete_likelihood}) is highly nonlinear in unknown parameters, we develop an iterative solution to the problem. We first formulate a set of Expectation-Maximization (EM) iterations \cite{Dempster77} and then involve a Newton's linearization to solve for the unknown parameters.

\subsection{Expectation Maximization Solution}
We select the pairs of random variables $(R_k,N_k)$, $k=1,2, \ldots, K$ as complete data. The complete data log-likelihood, denote it by $l_{cd}(\cdot)$, is given by
\begin{multline}
l_{cd}(\mathbf{R},\mathbf{N})=  - \frac{ 1 }{2\sigma^2 }\sum_{i=1}^{K}(R_i-G_i)^2 \\+{\text{terms not function of }}\mathbf{\theta}.
\end{multline}
The measurements $Z_i,$ $i=1,\ldots,K,$ form incomplete data. The mapping from complete data space to incomplete data space is given by $Z_k=q(R_k)+N_k$, where $q(.)$  is a known quantization function.

Denote by $\hat{\mathbf{\theta}}^{(k)}$ an estimate of the vector $\mathbf{\theta}$ obtained at the $k$-th iteration.  To update the estimates of parameters we alternate the expectation and maximization steps. During the expectation step, we evaluate the conditional expectation of the complete data log-likelihood:
\begin{eqnarray}
Q^{(k+1)}& = &E\left[ \left. - \frac{ 1 }{2\sigma^2 }\sum^{K}_{i =1}(R_i-G_i)^2 \right| {\mathbf{Z},\hat{\theta }^{(k)}}\right],
\label{eq:E-step}
\end{eqnarray}
where the expectation is with respect to the conditional probability density function of the complete data, given the incomplete data (measurements) and the estimates of the parameters at the $k$-th iteration. During the maximization step we maximize (\ref{eq:E-step}):
\begin{multline}
 \frac{dQ^{(k+1)}}{d\theta_t} = E\left[ \left. \frac{ 1 }{\sigma^2 }\sum^{K}_{i =1}(R_i-G_i)\frac{dG_i}{d\theta_t} \right| {\mathbf{Z},\hat{\theta }^{(k)}}\right]\Big|_{\hat{\theta }^{(k+1)}}
=0,\\\text{ $ t$}=1,\ldots,L.
\label{eq:M-step}
\end{multline}

To find the conditional expectation we note that the conditional probability density function (p.d.f.) of $Z_i,$  given $R_i,$  is Gaussian with mean $q(R_i)$  and variance $\eta^2$  and the p.d.f. of $R_i$ is Gaussian with mean $G_i$  and variance $\sigma^2.$  We also note that at the $k$-th iteration the conditional pdf of $R_i,$   given $Z_i,$   implicitly involves the estimates of the parameters obtained at the $k$-th iteration.

Denote by $G_i^{(k)}$ the estimate of the field $G(x,y)$ at the location $(x_i,y_i)$ with the vector of parameters $\mathbf{\theta}$ replaced by their estimates $\hat{\mathbf{\theta}}^{(k)}.$ Then
%
the final expression for the iterative evaluation of the unknown parameters can be written as
\begin{multline}
\sum^{K}_{i =1}\frac{dG_i^{(k+1)}}{d\theta_t}A(G_i^{(k)})-\sum^{K}_{i =1}G_i^{(k+1)}\frac{dG_i^{(k+1)}}{d\theta_t}B(G_i^{(k)})
=0, \\ t=1,\ldots,L,
\label{eq:structure}
\end{multline}
where
\begin{equation}
A(G_i^{(k)})=\sum_{j=1}^M \frac{\exp\left(-\frac{(z_i-\nu_j)^2}{2\eta^2}\right)}{f_{Z_i}^{(k)}(z_i)\sqrt{2 \pi \eta^2}} \left( \sqrt{\frac{\sigma^2}{2\pi}} e^{-\frac{(\tau_j-G_i^{(k)})^2}{2\sigma^2}} \right. \end{equation}
\[ \left. - \sqrt{\frac{\sigma^2}{2\pi}} e^{-\frac{(\tau_{j+1}-G_i^{(k)})^2}{2\sigma^2}} + G_i^{(k)} \Delta Q^{(k)}(j,i)\right), \]
\begin{equation}
B(G_i^{(k)})=\sum^{M}_{j =1}\frac{ \exp \left(-\frac{(z_i-\nu_j)^2 }{2\eta^2 }\right)}{f_{Z_i}^{(k)}(z_i)\sqrt{2\pi\eta^2}}\Delta Q^{(k)}(j,i),
\end{equation}
with $\Delta Q^{(k)}(j,i)= Q\left(\frac{\tau_j-G_i^{(k)}}{\sigma}\right)-Q\left(\frac{\tau_{j+1}-G_i^{(k)}}{\sigma}\right)$ and \[f_{Z_i}^{(k)}(z_i)=\int f^{(k)}(z_i|r)f^{(k)}(r)dr.\]
The expression $Q(\cdot)$ is used to denote the Q-function.


\subsection{Linearization}
\label{sec:Linearization}
The equations (\ref{eq:structure}) are nonlinear in $\hat{\mathbf{\theta}}^{(k+1)}$ and have to be solved numerically for each iteration. To simplify the solution, we linearize the expression in (\ref{eq:structure}) by means of Newton's method. Denote by $\mathbf{F}(\theta^{(k+1)})$ the vector form of the left side in (\ref{eq:structure}), which is a mapping from $\mathbf{\theta}^{(k+1)}$ to the range of $\mathbf{F}(\theta^{(k+1)})$. Let $J\left(\theta_n^{(k+1)}\right)$ be the Jacobian of the mapping. The index $n$ indicates the iteration of the Newton's solution. Then $\mathbf{\theta}^{(k)}$ solves the following linearized equation:
\begin{eqnarray} \label{}
J\left(\theta_n^{(k+1)}\right)\left(\theta_{n+1}^{(k+1)}-\theta_{n}^{(k+1)}\right)=-F \left( \theta_n^{(k+1)}\right).
\end{eqnarray}


\section{Numerical Analysis}
\label{sec:Numerical_Analysis}
\begin{figure}[!t]
\centering
\includegraphics[width=0.9\linewidth]{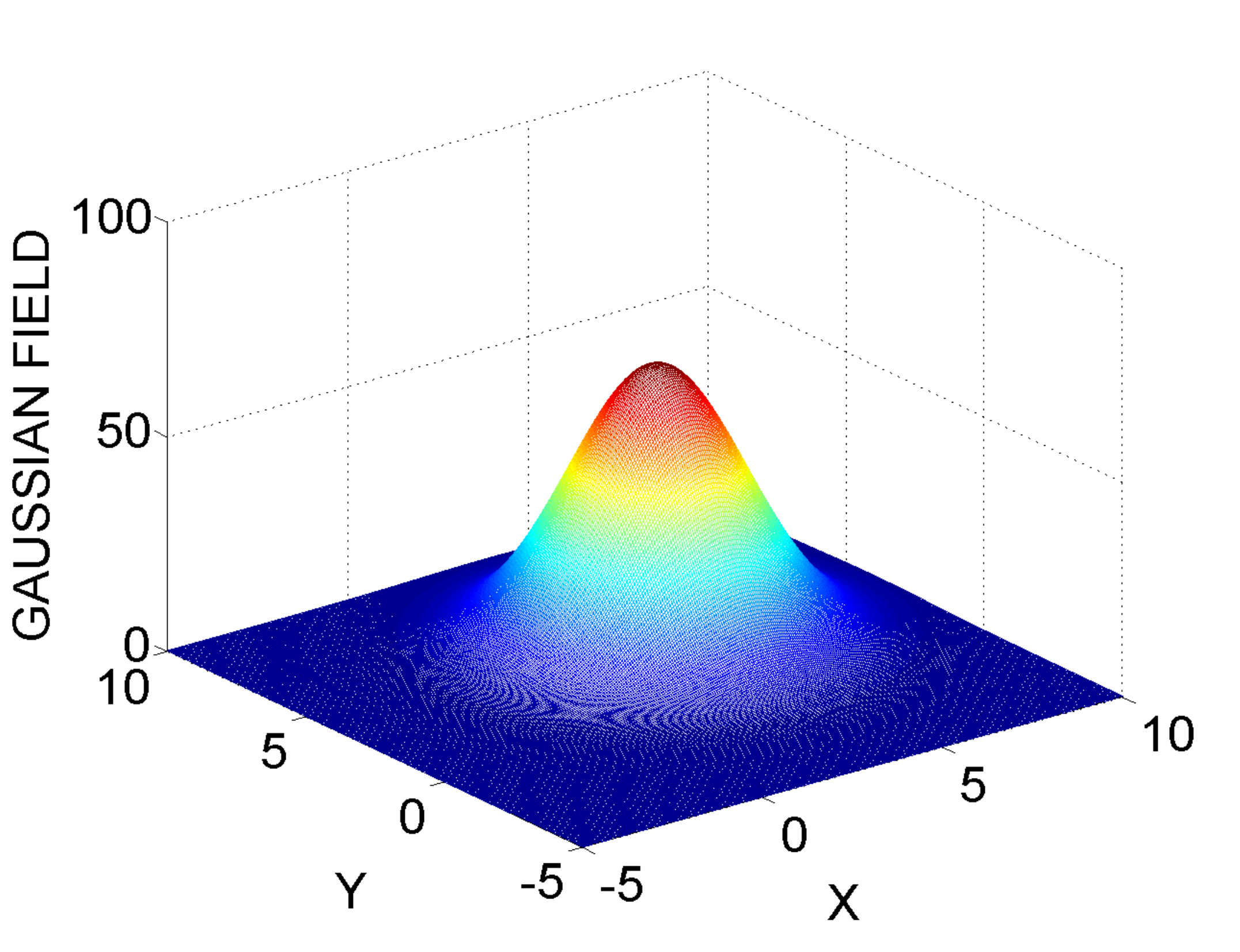}
\caption{Squared Gaussian field located at $(x_c,y_c)=(4,4)$ with peak parameter $4$ and variance $4.$ }
\label{fig:Gaussian_field}
\end{figure}

In this section, the performance of the distributed ML estimator in (\ref{eq:structure}) is demonstrated on simulated data. A distributed network of $K$ sensors is formed by positioning sensors at random over an area $A$ of size $8 \times 8.$ The location of each sensor is noted. A Gaussian field shown in Fig. \ref{fig:Gaussian_field} is sampled at the location of the $i$-th sensor, $i=1,\ldots,K$ and a sample of randomly generated Gaussian noise with mean zero and variance $\sigma^2$ is added to each field measurement. In our simulations, $K$ is varied from $5$ to $200$ and $\sigma^2$ is selected such that the total signal-to-noise ratio (SNR) of the local observations defined as
\begin{equation}
SNR_O = \frac{ {\int \int}_A G^2(x,y:\theta)dxdy}{A \sigma^2}
\end{equation}
is $15$ dB. Each sensor observation is quantized to one of $M$ levels using a  uniform deterministic quantizer. We set the number of quantization levels to $M=8$ and the quantization step to $8.$  $K$ parallel white Gaussian noise channels add samples of noise with variance $\eta^2$  selected to set the total SNR during data transmission defined as
\begin{equation}
SNR_C = \frac{ {\int \int}_A E\left[q^2(R(x,y:\theta))\right]dxdy}{A \eta^2}
\label{eq:SNR_C}
\end{equation}
to $15$ dB, and the FC observes the noisy quantized samples of the field. The function $q(R(x,y:\theta))$ in (\ref{eq:SNR_C}) is a quantized version of $R(x,y:\theta)=G(x,y:\theta)+W.$

\begin{figure}[!t]
\centering
\includegraphics[width=0.9\linewidth]{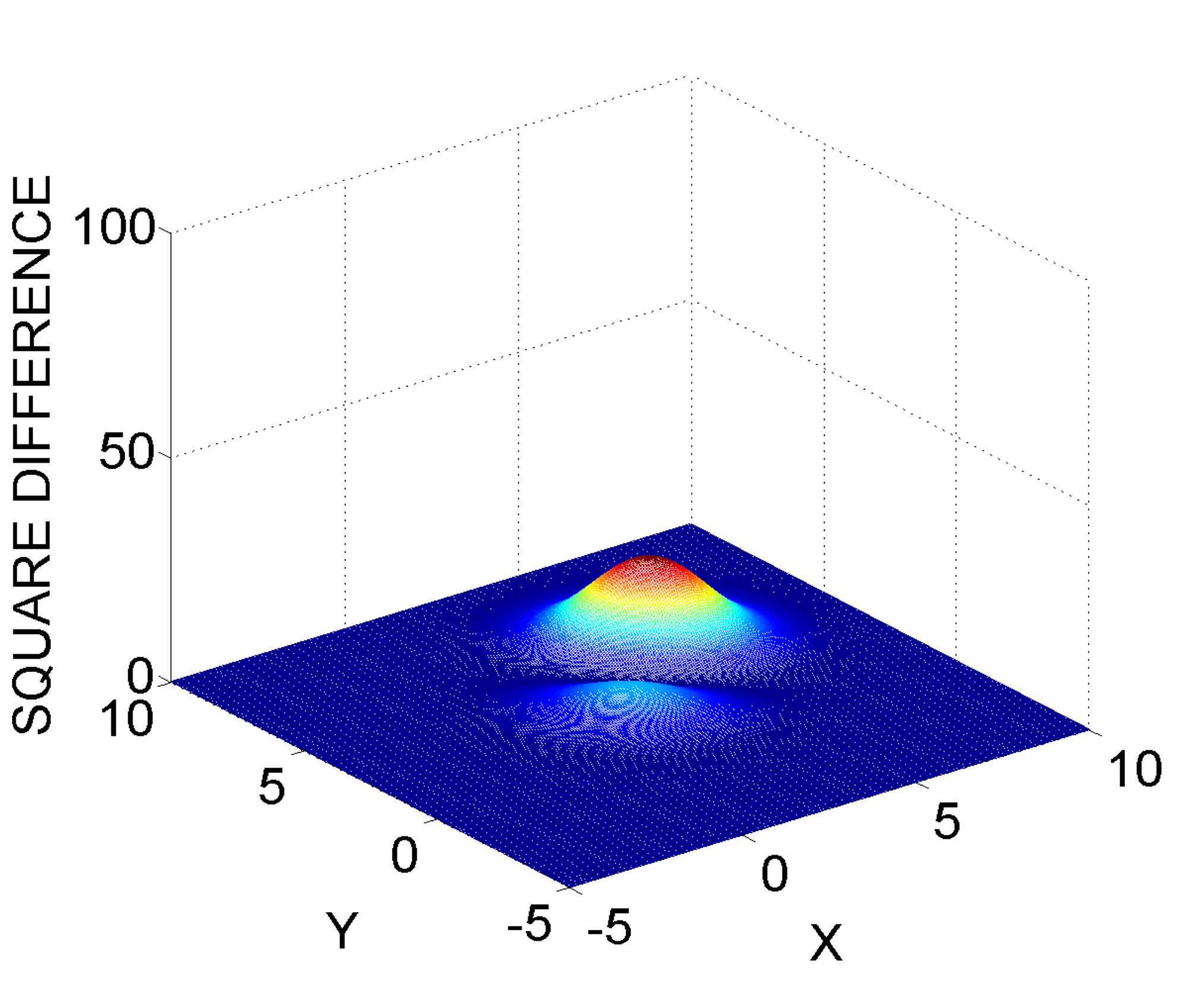}
\caption{The squared difference between the original and reconstructed fields. }
\label{fig:Difference_field}
\end{figure}

%

\begin{figure}[!t]
\centering
\includegraphics[width=1.9in]{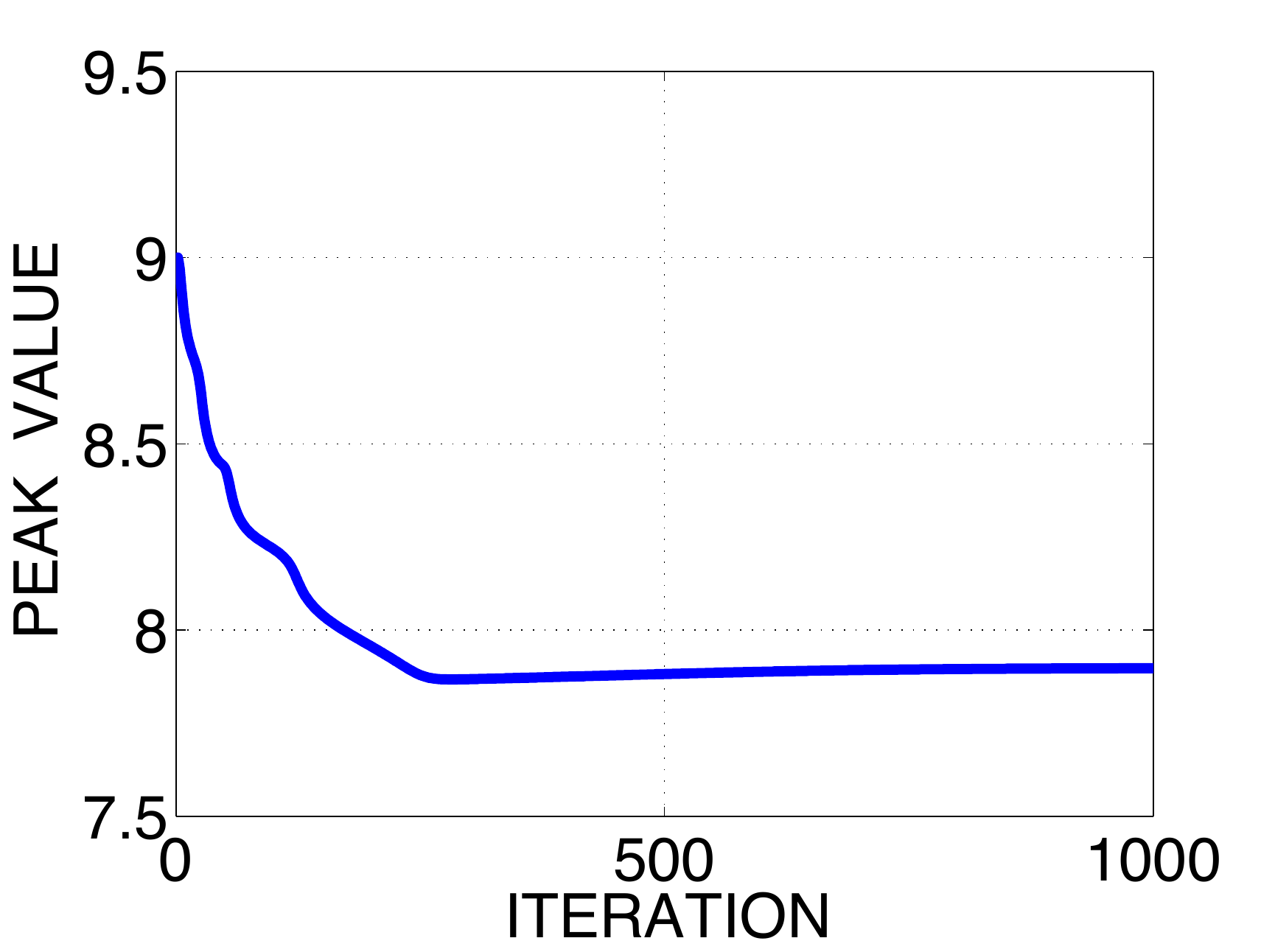}
\caption{Illustration of the EM convergence for $M=8.$ Peak parameter.}
\label{fig:Peak_parameter}
\end{figure}

\begin{figure}[!t]
\centering
\includegraphics[width=1.9in]{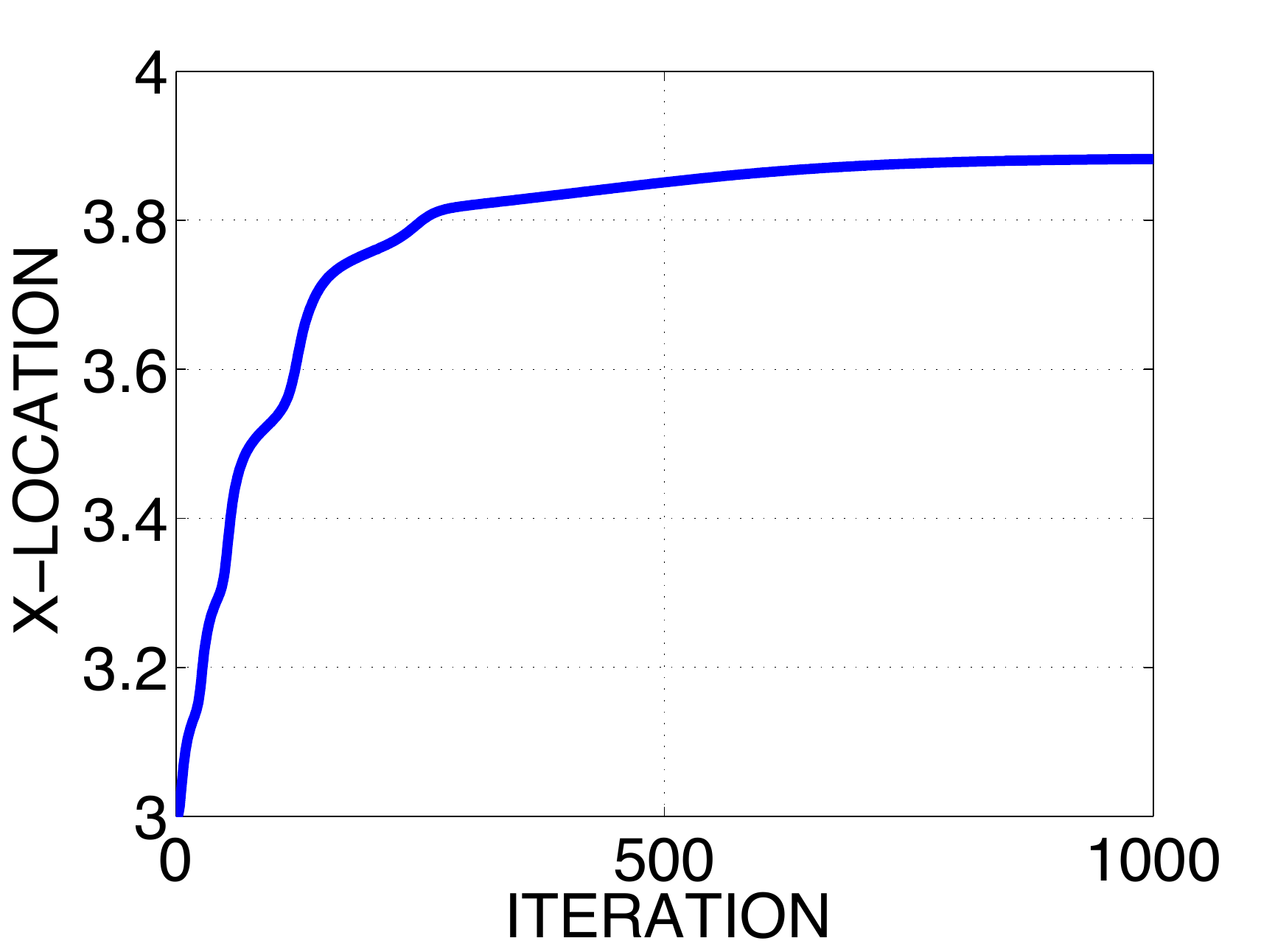}
\caption{ Illustration of the EM convergence for $M=8.$ X-location.}
\label{fig:X_location}
\end{figure}

\begin{figure}[!t]
\centering
\includegraphics[width=1.9in]{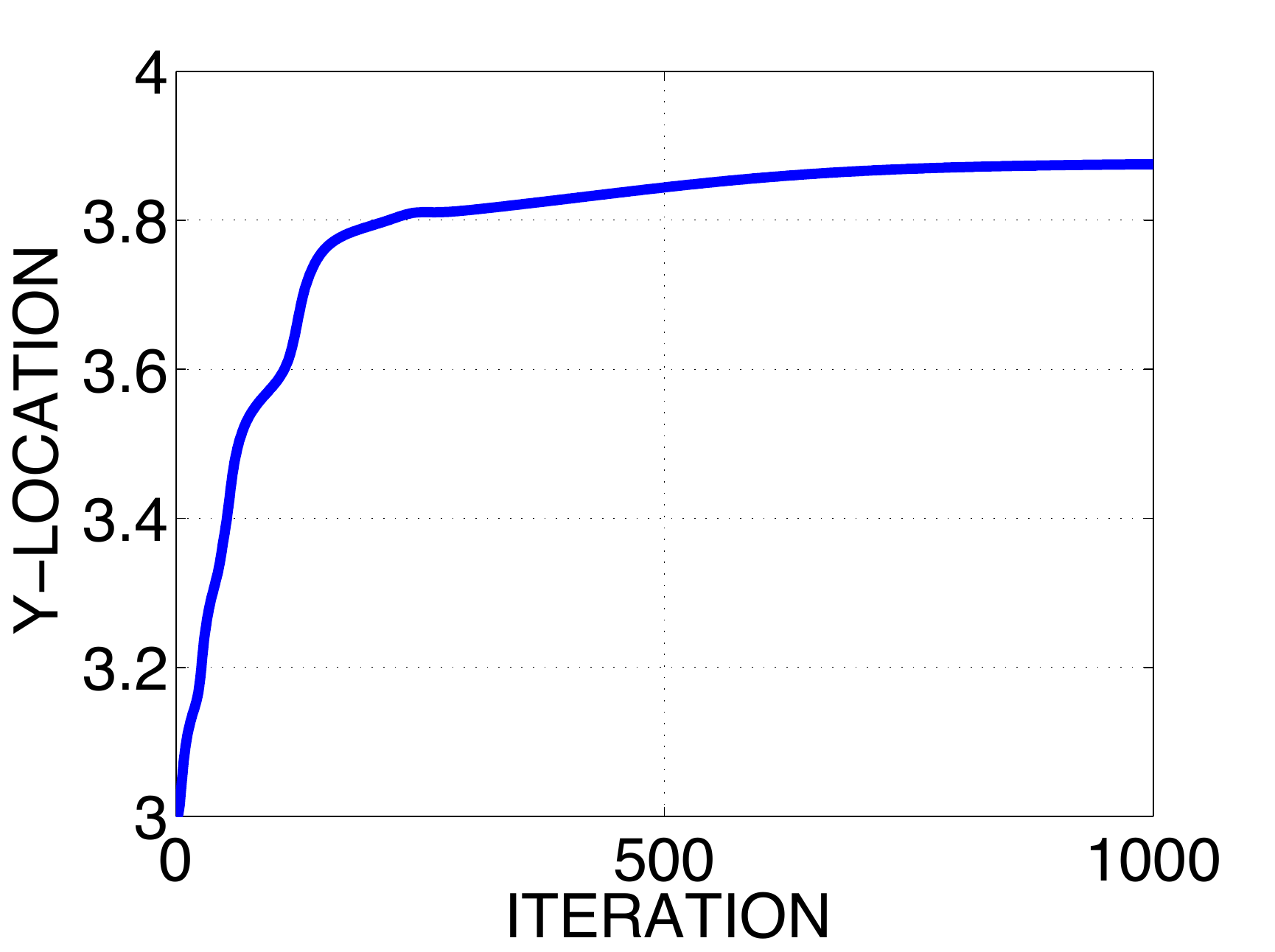}
\caption{ Illustration of the EM convergence for $M=8.$ Y-location.}
\label{fig:Y_location}
\end{figure}

First, we illustrate the convergence of the EM algorithm. The value of the ML estimate as a function of iteration is shown in Figs. \ref{fig:Peak_parameter}, \ref{fig:X_location} and  \ref{fig:Y_location} for the peak value of the field, its $x$-location and its $y$-location, respectively. This illustration is based on a single realization of the distributed network with $K=10$ and $M=8.$ We can observe that with the initial values $9$ for the peak of the field, $3$ for the x-location and $3$ for the y-location, the algorithm takes about $600$ EM iterations to converge to the final values $7.90,$ $3.88,$ and $3.88,$ respectively. The true values of these parameters are $8,$ $4,$ and $4.$ The discrepancy between the vectors of estimated and true parameters are due to a low sensor density in the network, relatively rough quantization, and the distortions due to sensor and channel noise.

The square distance per pixel between the original and reconstructed Gaussian fields is displayed in Fig. \ref{fig:Difference_field}.

To further analyze the estimation performance, we evaluate the mean square error (MSE) between the estimated and true location parameters. The MSE is evaluated numerically by means of 1000 Monte Carlo simulations. Each vector of estimated parameters is substituted back in the expression for the parametric field, and an integrated square error (ISE) between the true and estimated fields is evaluated. The integrated square error (ISE) is defined as

\begin{equation}
ISE= \frac{{\int \int}_A  |\hat{G}(x,y)-G(x,y)|^2 dxdy}{{\int \int}_A  |G(x,y)|^2 dxdy}.
\end{equation}
The ISE statistically averaged over 1000 Monte Carlo simulations is an approximation to integrated mean square error (IMSE).
\begin{figure}[!t]
\begin{center}
   \includegraphics[width=0.9\linewidth]{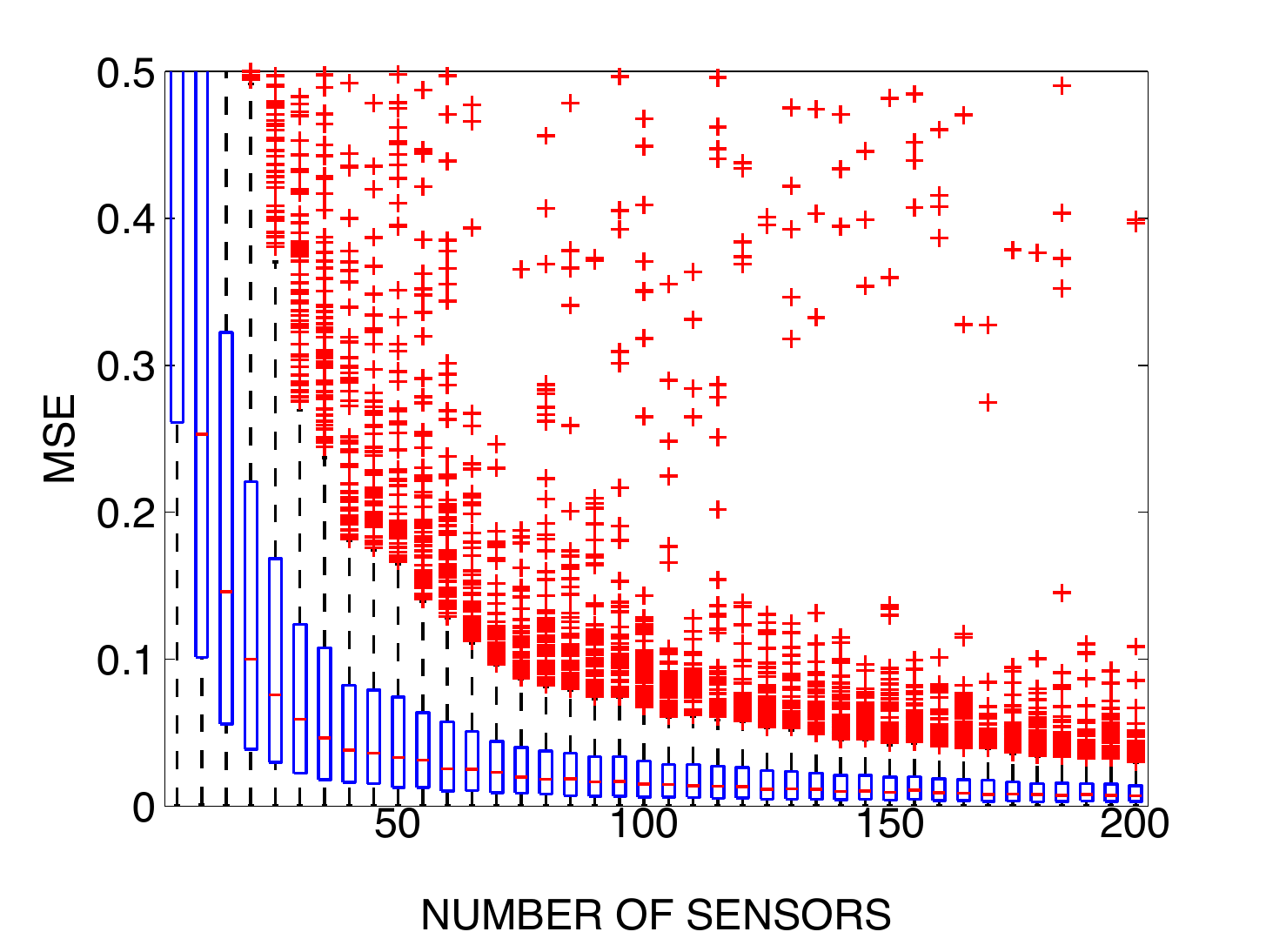}
\end{center}
 \vspace{-0.5cm}
   \caption{A box plot of the SE between the estimated and true location of the object displayed as a function of the number of sensors distributed over the area $A.$ The number of quantization levels is set to $M=8.$}
\label{fig:MSE_box_plot}
\end{figure}

\begin{figure}[!t]
\begin{center}
   \includegraphics[width=0.9\linewidth]{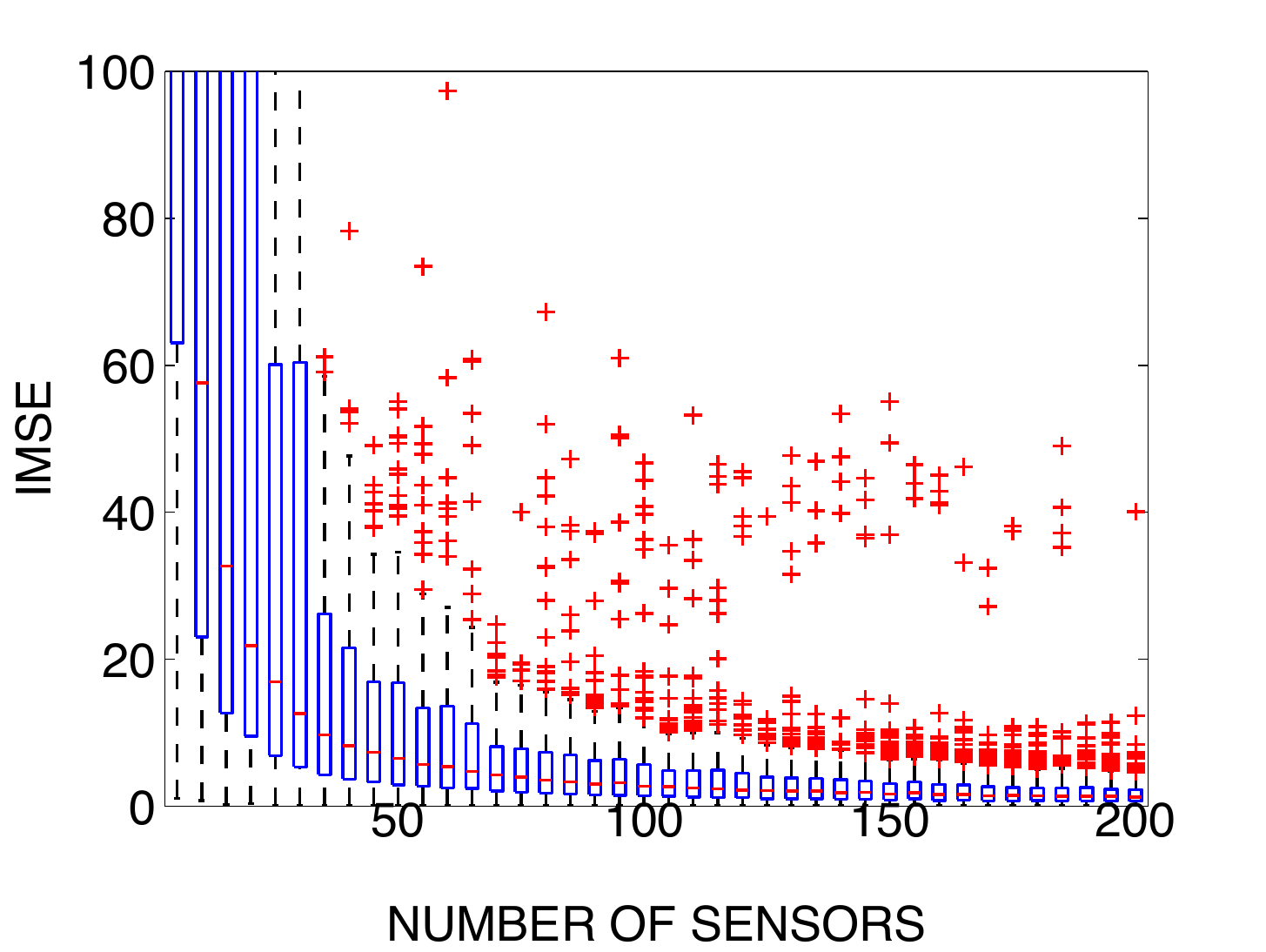}
\end{center}
 \vspace{-0.5cm}
   \caption{Dependence of the simulated ISE on the number of sensors distributed over the area $A$ for the case of $M=8.$}
\label{fig:box_plot_quantized}
\end{figure}
\begin{figure}[!t]
\begin{center}
   \includegraphics[width=0.9\linewidth]{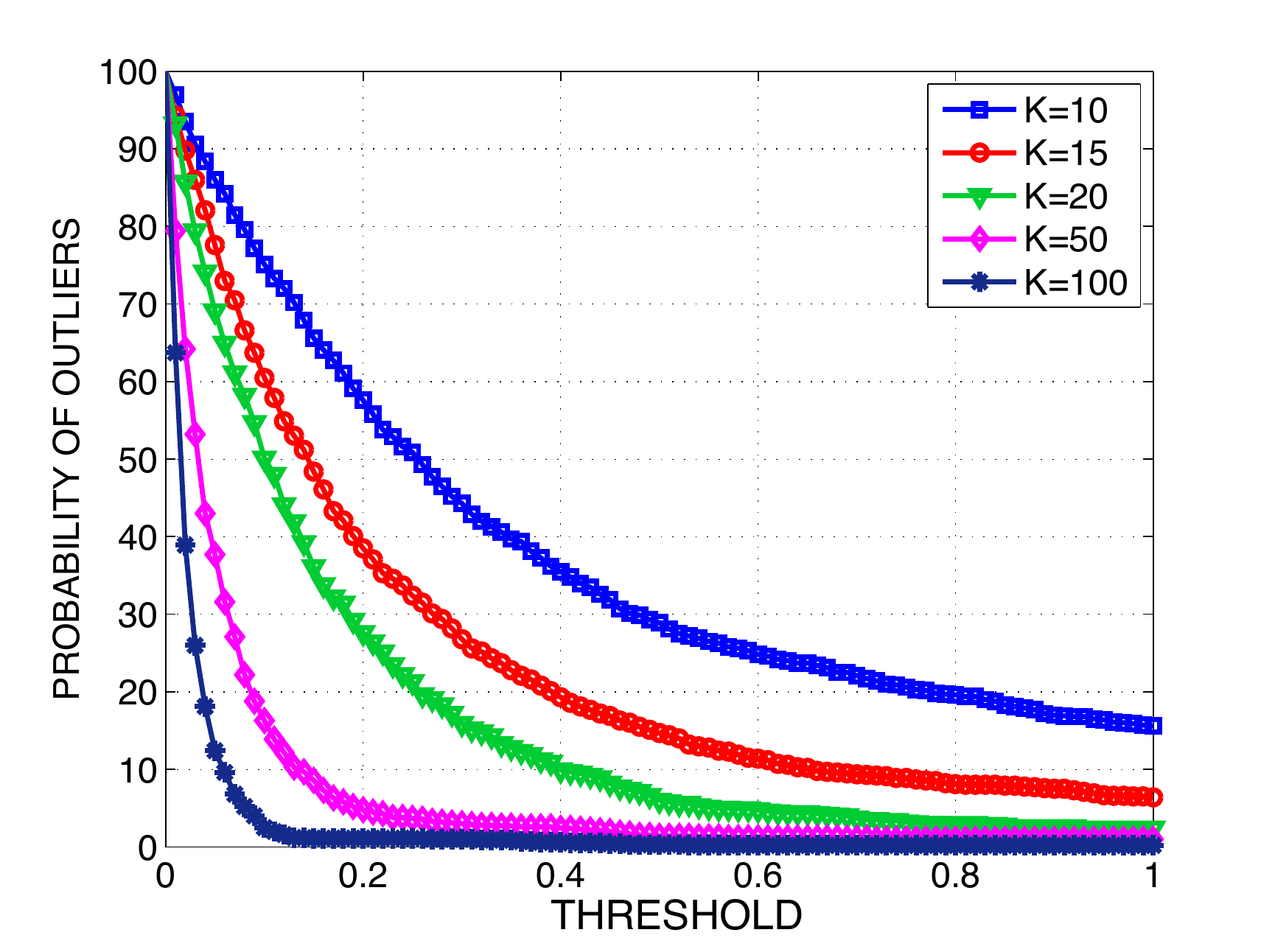}
\end{center}
 \vspace{-0.5cm}
   \caption{Probability of outliers $P_{outliers}(\tau)=P[SE > \tau]$ (expressed in percents) as a function of $\tau.$ The plot is based on $1000$ Monte Carlo simulations.}
\label{fig:P_outliers_MSE}
\end{figure}

The dependence of the MSE on the number of sensors, $K,$ in the distributed network for the case of $M=8$ quantization levels is shown in Fig. \ref{fig:MSE_box_plot}.
The dependence of the IMSE on the number of sensors (sensor density) in the distributed network for the same value of $M$ is displayed in Fig. \ref{fig:box_plot_quantized}. The number of sensors distributed over the area $A$ is varied from $5$ to $200$ with the step $5.$  Each box in Fig. \ref{fig:MSE_box_plot} and Fig. \ref{fig:box_plot_quantized} is generated using $1000$ Monte Carlo realizations of the network and EM runs. The central mark in each box is the median. The edges of the box present the $25$th and $75$th percentiles. The dashed vertical lines mark the data that extend beyond the two percentiles, but not considered as outliers. The outliers are plotted individually and marked with a ``+'' sign. The percentage of outliers due to divergence of the EM algorithm is depicted in Fig. \ref{fig:P_outliers_MSE}. Note the large percentage of outliers for small values of $K,$ $K=10, 15, 20.$  These correspond to the case when one of the three parameters did not converge to its true value.

The results indicate that the location estimation and the field reconstruction of a relatively good quality is possible with $M=8$ and the number of sensors equal or exceeding $20.$

\begin{figure}[!t]
\begin{center}
   \includegraphics[width=0.9\linewidth]{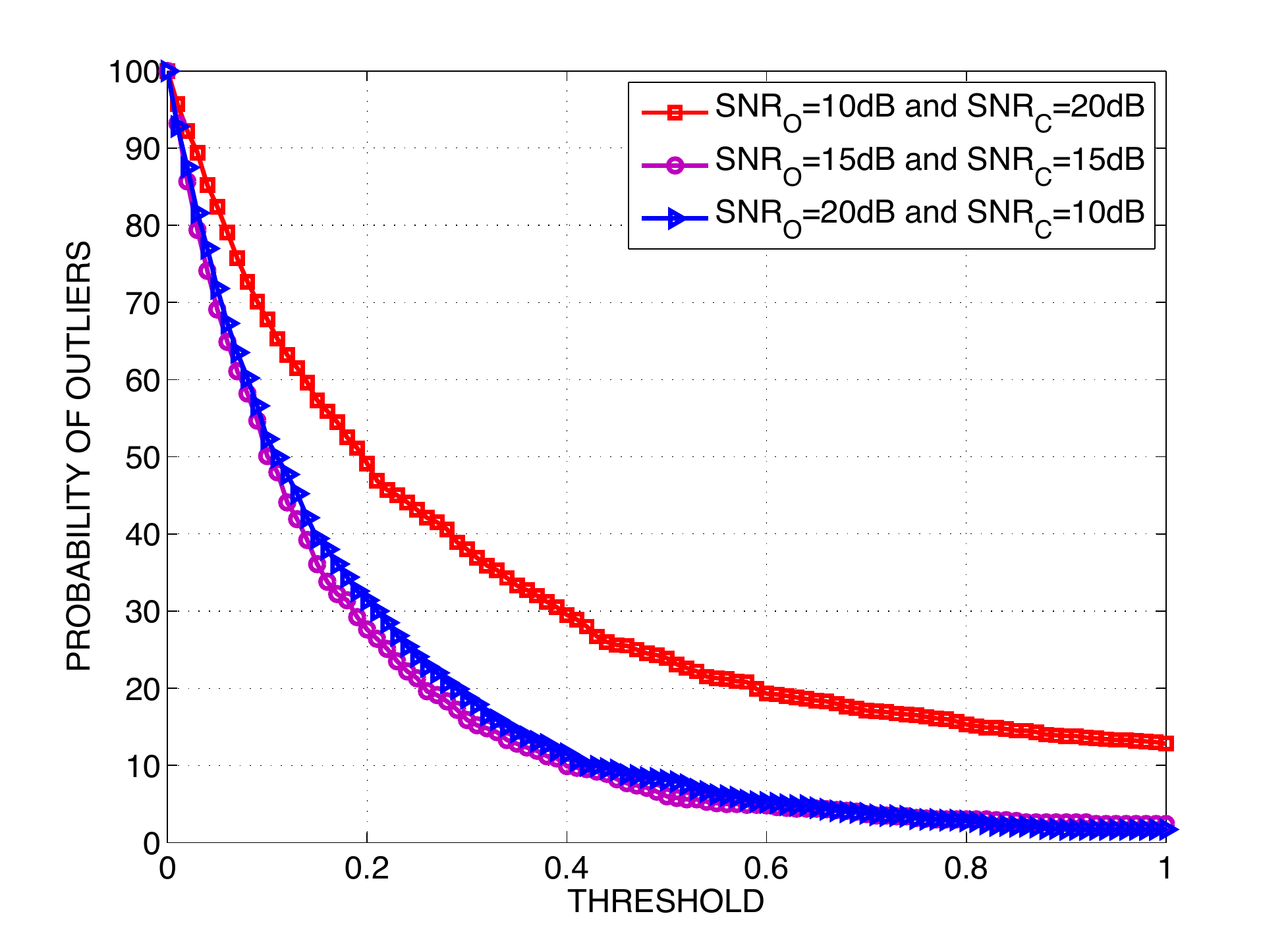}
\end{center}
 \vspace{-0.5cm}
   \caption{Probability of outliers (expressed in percents) as a function of the threshold for different values of SNR in observation and transmission channels.}
\label{fig:P_outliers_SNR}
\end{figure}
Fig. \ref{fig:P_outliers_SNR} compares the percentage of outliers plotted as a function of varying threshold for three different realizations of $SNR_O$ and $SNR_C.$  Note that for $M=8$ the effect of the SNR in the observation channel is more pronounced compared to the SNR in the transmission channel. The case of high $SNR_O=20$ dB and low $SNR_C=10$ dB is preferred by the estimator compared to the case of low $SNR_O=10$ dB and high $SNR_C=20$ dB.

\begin{figure}[!t]
\begin{center}
   \includegraphics[width=0.9\linewidth]{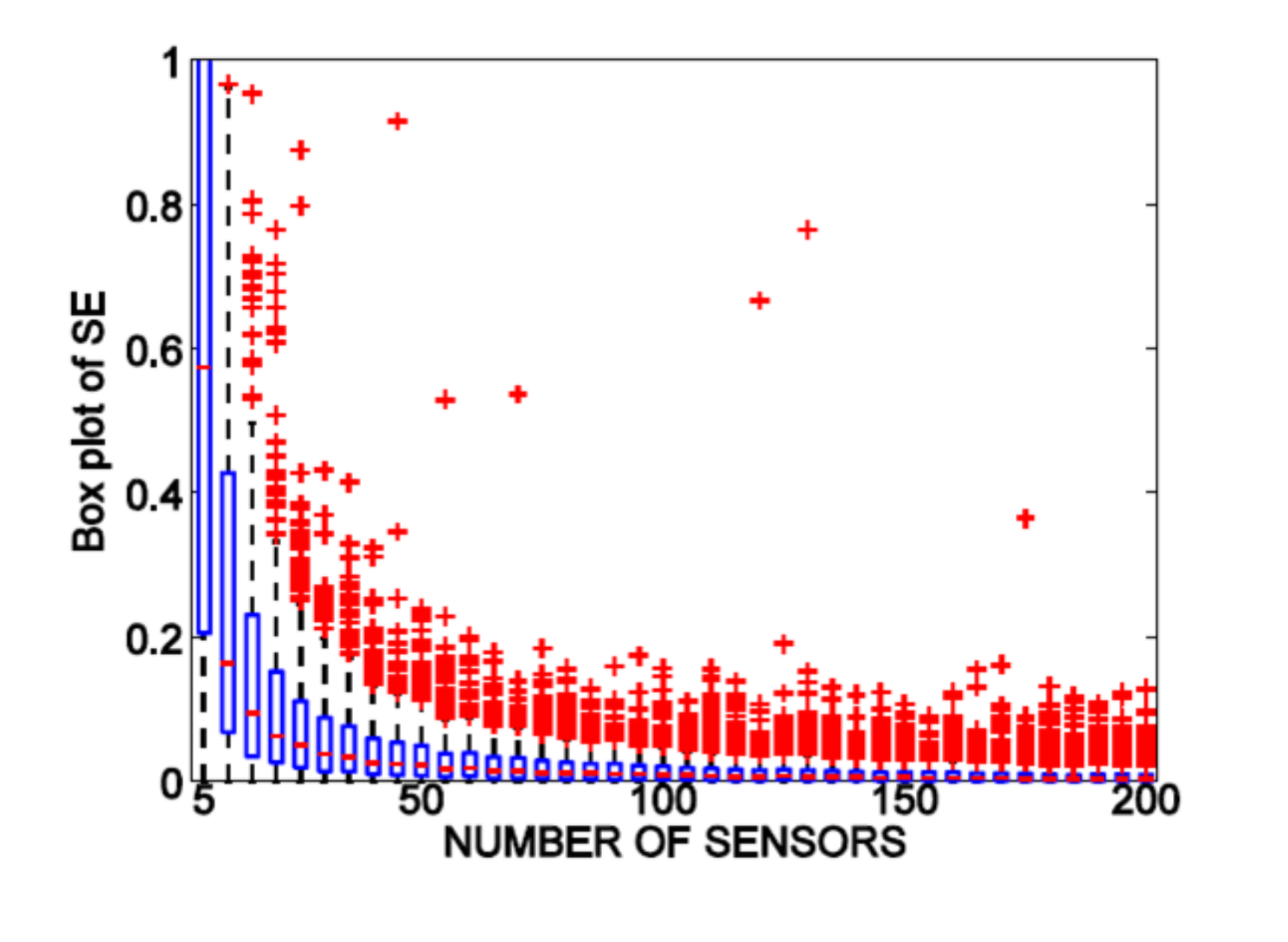}
\end{center}
\caption{A box plot of the SE between the estimated and true location of the object displayed as a function of the number of sensors distributed over the area $A.$ The number of quantization levels is set to $M=16.$}
\label{fig:SE_M16}
\end{figure} 

\begin{figure}[!t]
\begin{center}
   \includegraphics[width=0.9\linewidth]{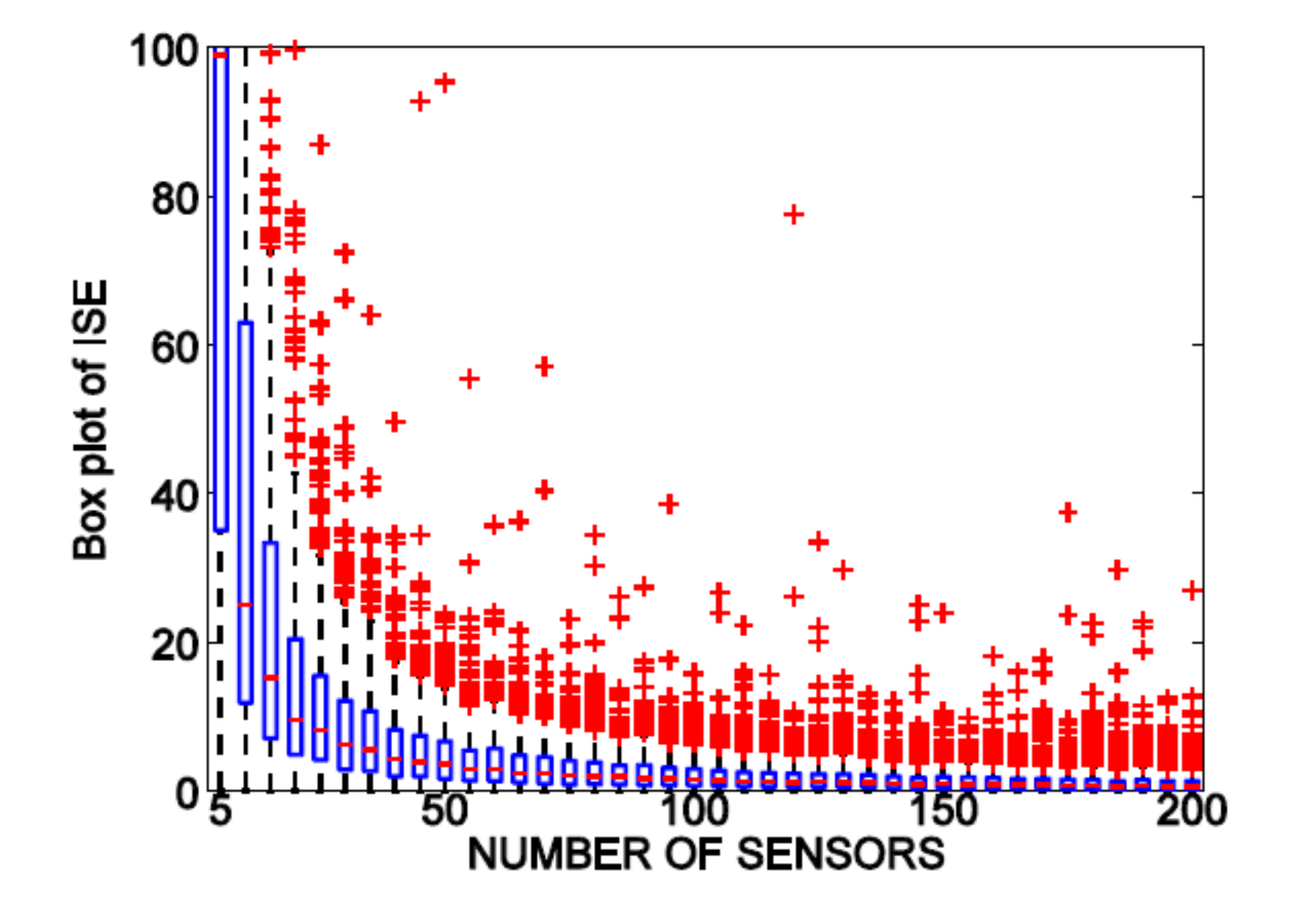}
\end{center}
\caption{Dependence of the simulated ISE on the number of sensors distributed over the area $A.$ The number of quantization levels is set to $M=16.$}
\label{fig:ISE_M16}
\end{figure}
   
\begin{figure}[!t]
\begin{center}
   \includegraphics[width=0.9\linewidth]{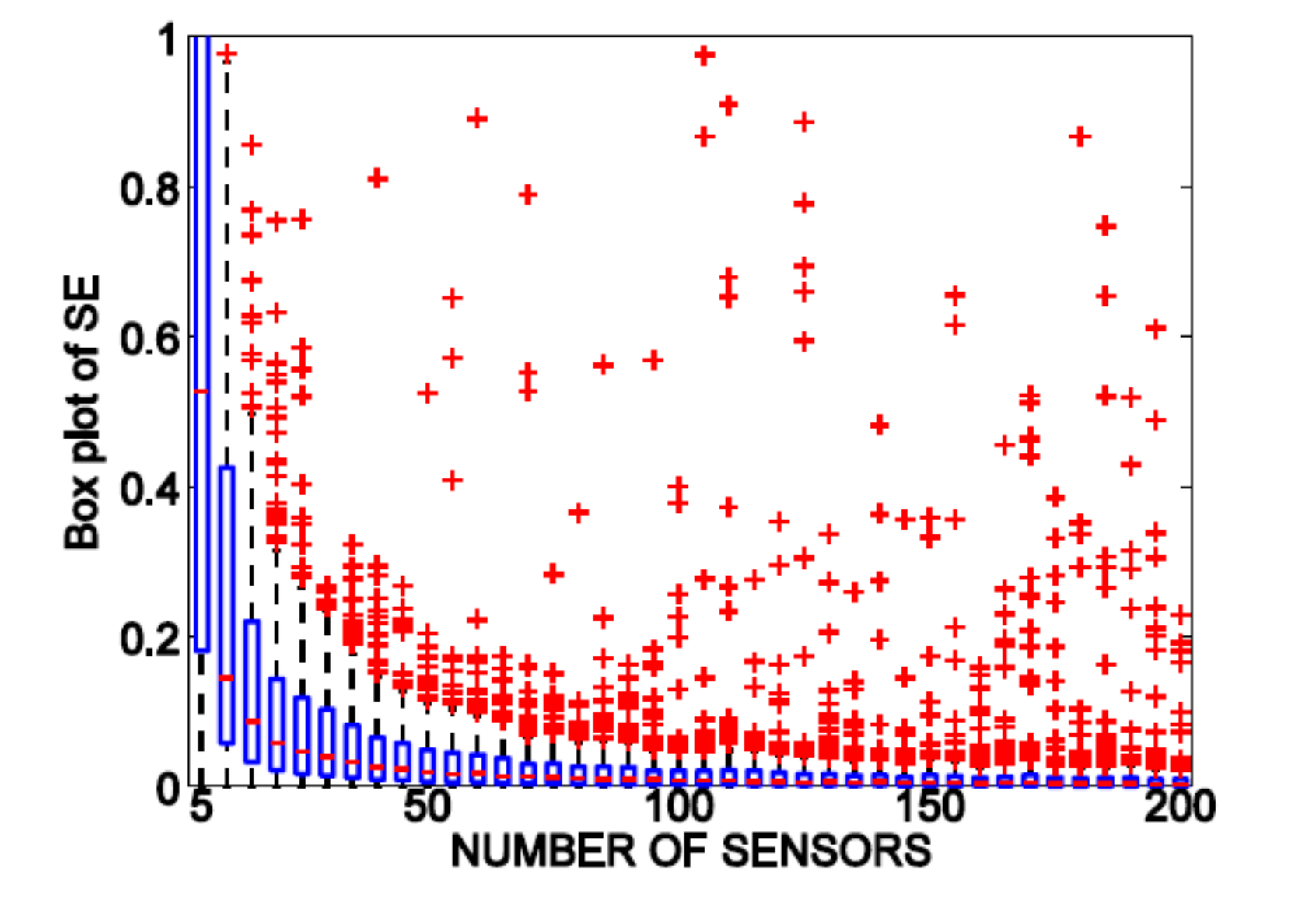}
\end{center}
\caption{A box plot of the SE between the estimated and true location of the object displayed as a function of the number of sensors distributed over the area $A.$ The number of quantization levels is set to $M=32.$}
\label{fig:SE_M32}
\end{figure}

\begin{figure}[!t]
\begin{center}
   \includegraphics[width=0.9\linewidth]{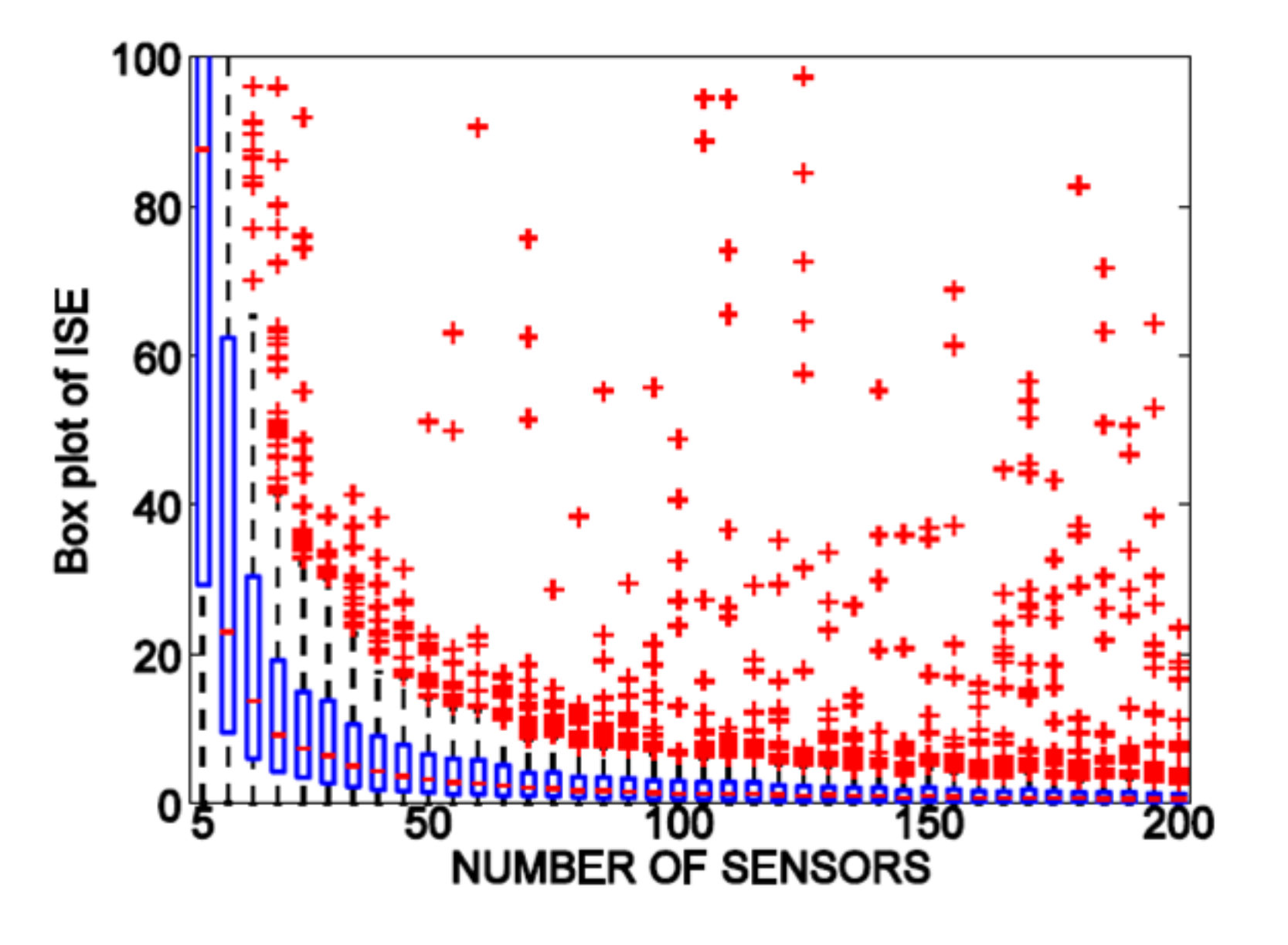}
\end{center}
\caption{Dependence of the simulated ISE on the number of sensors distributed over the area $A.$ The number of quantization levels is set to $M=32.$}
\label{fig:ISE_M32}
\end{figure}

%

%
%
%
%
%
%
%

A set of box plots showing dependence of the SE and the ISE on the number of sensors distributed over the area $A$ for $M=16$ and $M=32$ have been also generated. The results are similar to those for the case of $M=8$ with the difference that the number of outliers as a function of the threshold decays faster to zero (see Figs. \ref{fig:SE_M16}, \ref{fig:ISE_M16}, \ref{fig:SE_M32}, and \ref{fig:ISE_M32}) for illustration).


%
%
%



\section{Summary}
\label{sec:Summary}

In this paper, a distributed ML estimation procedure for estimating a parametric physical field is formulated. An iterative linearized EM solution is presented and numerically evaluated. The model of the network assumed (1) independent Gaussian sensor and transmission noise; (2) quantization of sensory data prior to transmission; and (3) parametric function estimation at the FC. The stability of the EM algorithm has been evaluated for three different values of $SNR_O$ and $SNR_C.$ The results show that for a small number of quantization levels (quantization error is large) $SNR_O$ dominates $SNR_C$ in terms of its effect on the performance of the estimator. Also, when the sensor network is sparse, $K=10, 15, 20$  the EM algorithm produces a substantial number of outliers. Denser networks, $K>20,$ are more stable in terms of reliable parameter estimation.  A similar analysis has been performed for $M=16$ and $M=32.$

In the future, we plan to analyze the estimation abilities of the network at low SNR values and develop a Cramer-Rao bound on the estimated parameters.


%

\appendices
\section{}
This section provides details leading to the equation (\ref{eq:structure}). Consider the $i$-th term under the sum in (\ref{eq:M-step}):
\[ E\left[ \left. (r_i-G_i)\frac{\partial G_i}{\partial \theta_t} \right| z_i,\hat{\theta}^{(k)} \right]   \]
\[ = \int_{-\infty}^{+\infty} (r_i-G_i) \frac{\partial G_i}{\partial \theta_t} \frac{\exp\left( -\frac{(r_i-G_i^{(k)})^2}{2\sigma^2}\right)}{f_{Z_i}^{(k)}(z_i) \sqrt{2\pi\sigma^2}} \]
\[ \times \frac{\exp\left(-\frac{(z_i-q^{(k)}(r_i))^2}{2 \eta^2}\right)}{\sqrt{2\pi \eta^2}} dr_i \]
\[ = \sum_{j=1}^M \int_{\tau_j}^{\tau_{j+1}} (r_i-G_i) \frac{\partial G_i}{\partial \theta_t} \frac{\exp\left(-\frac{(r_i-G_i^{(k)})^2}{2\sigma^2}\right)}{f_{Z_i}^{(k)}(z_i)  \sqrt{2\pi\sigma^2}} \]
\[ \times \frac{\exp\left(-\frac{(z_i-\nu_j)^2}{2 \eta^2}\right)}{\sqrt{2\pi \eta^2}} dr_i \]
\[ = \sum_{j=1}^M \frac{\exp\left(-\frac{(z_i-\nu_j)^2}{2 \eta^2}\right)}{f_{Z_i}^{(k)}(z_i) \sqrt{2\pi \eta^2}} \frac{\partial G_i}{\partial \theta_t} \]
\[ \times \int_{\tau_j}^{\tau_{j+1}} (r_i-G_i)\frac{\exp\left(-\frac{(r_i-G_i^{(k)})^2}{2\sigma^2}\right)}{\sqrt{2\pi\sigma^2}} dr_i.\]

Note that the difference $(r_i-G_i)$ in the last integral can be rewritten as $(r_i-G_i^{(k)}+G_i^{(k)}-G_i).$ Then
\[E\left[ \left. (r_i-G_i)\frac{\partial G_i}{\partial \theta_t} \right| z_i,\hat{\theta}^{(k)} \right] = \sum_{j=1}^M \frac{\exp\left(-\frac{(z_i-\nu_j)^2}{2\eta^2}\right)}{f_{Z_i}^{(k)}(z_i) \sqrt{2\pi\eta^2}} \frac{\partial G_i}{\partial \theta_t} \]
\[ \times \left\{ \frac{1}{\sqrt{2 \pi \sigma^2}} \int_{\tau_j}^{\tau_{j+1}} \exp \left( -\frac{(r_i-G_i^{(k)})^2}{2\sigma^2} \right) d \frac{(r_i-G^{(k)}_i)^2}{2} \right. \]
\[ \left. + (G^{(k)}_i-G_i) \frac{1}{\sqrt{2\pi \sigma^2}} \int_{\tau_j}^{\tau_{j+1}} \exp \left(-\frac{(r_i-G_i^{(k)})^2}{2 \sigma^2}\right) dr_i \right\}. \]

Replacing the last integral with a difference of two Q-functions $Q\left(\frac{\tau_j-G_i^{(k)}}{\sigma}\right)$ and $Q\left(\frac{\tau_{j+1}-G_i^{(k)}}{\sigma}\right)$ we obtain:
\[ \sum_{i=1}^K E\left[ \left. (r_i-G_i)\frac{\partial G_i}{\partial \theta_t} \right| z_i,\hat{\theta}^{(k)} \right] = \sum_{i=1}^K \sum_{j=1}^{M} \frac{\exp\left(-\frac{(z_i-\nu_j)^2}{2\eta^2}\right)}{f_{Z_i}^{(k)}(z_i) \sqrt{2\pi\eta^2}} \frac{\partial G_i}{\partial \theta_t} \]
\[\times \left\{ \frac{\sigma^2}{\sqrt{2 \pi \sigma^2}} \left\{ \exp \left( -\frac{(\tau_j-G_i^{(k)})^2}{2\sigma^2}\right) - \exp \left(-\frac{(\tau_{j+1}-G_i^{(k)})^2}{2\sigma^2}\right) \right\} \right. \]
\[ \left. + (G_i^{(k)}-G_i) \left\{ Q\left(\frac{\tau_j-G_i^{(k)}}{\sigma}\right) \right. \right. \]
\[ \left. \left. \left.- Q\left(\frac{\tau_{j+1}-G_i^{(k)}}{\sigma}\right) \right\} \right\} \right|_{G_i=G_i^{(k+1)}} = 0.\]

%
%
%
%
%
%

\ifCLASSOPTIONcaptionsoff
  \newpage
\fi

\balance



%

\end{document}